# Excitonic effects on infrared vibrational and Raman spectroscopy from first principles

Yang-Hao Chan,[1, 2] Zhenglu Li,[3, 4, 5] and Steven G. Louie[4, 5]

[1]*Institute of Atomic and Molecular Sciences, Academia Sinica, Taipei 10617, Taiwan*
[2]*Physics Division, National Center of Theoretical Sciences, Taipei 10617, Taiwan*
[3]*Mork Family Department of Chemical Engineering and Materials Science,*
*University of Southern California, Los Angeles, California 90089, USA*
[4]*Department of Physics, University of California at Berkeley, Berkeley, California, 94720, USA*
[5]*Materials Sciences Division, Lawrence Berkeley National Laboratory, Berkeley, California 94720, USA*
(Dated: June 22, 2024)

We develop a first-principles approach to compute infrared (IR) vibrational absorption and Raman scattering spectra with excitonic effects included. Our method is based on a perturbative expansion of electron-phonon and electron-light couplings in the time-dependent adiabatic GW (TD-aGW) theory. We show that excitonic effects in the IR absorption spectrum can be included by replacing the free electron-hole propagators in the perturbative expression for independent particles with their interacting counterparts, which are readily available from standard GW-Bethe-Salpeter equation calculations. For Raman spectrum, our derived expression agrees with the single and double resonance terms from a diagrammatic approach. We show significant excitonic enhancement in both the IR and resonance Raman scattering intensity for monolayer $MoS_2$, $WS_2$, and $WSe_2$. Moreover, the exciton-phonon coupling strength and exciton energy landscape can be accessed by analyzing resonance Raman spectrum of these materials.

## I. INTRODUCTION

Infrared (IR) absorption spectrum and Raman scattering [1–3] are two important optical spectroscopy methods to analyze the microscopic electron and phonon structures and light-matter interactions in solids. In the former, light frequency is resonant with the energy of the vibrational modes, whereas in the latter, a shift in energy of the scattering light is detected. Combined with first-principles calculations, IR and Raman spectroscopies have been widely used to characterize sample qualities, strain effects, crystal symmetries, and twisting angles in stacked two-dimensional (2D) materials, among many other applications. In both spectra, in the lowest order electron-phonon coupling, energies of isolated peaks are identified as zone center phonon excitations and the spectral intensity is connected to the electron-phonon coupling strength. The importance of excitons and exciton-phonon coupling are also becoming more recognized as the study of quasi-low dimensional quantum materials draws significant attention recently.

Excitons are correlated electron-hole excited states (with most prominent ones being bound electron-hole pairs with energy in the gap of an insulator) and are fundamental excitations in optical responses. Excitons dominate optical absorption and photoluminescence spectra especially in low dimensional materials [4] due to reduced screening and thus enhanced Coulomb interaction. Nowadays, first-principles calculations of linear optical absorption spectra with excitonic effects included are routinely performed with the GW-Bethe-Salpeter equation (BSE) method [5–7] and the results are typically in excellent agreement with experiments for many materials. In contrast, excitonic effects in IR absorption and Raman scattering are less investigated. Recent experimental results have shown that Raman spectra intensity as a function of laser frequency (resonance Raman spectroscopy) can reveal information of exciton states and exciton-phonon coupling [8–12]. Moreover, higher-order Raman spectra have been used to detect exciton scattering pathways, providing valuable information on exciton dynamics [13]. To better interpret experiments, accurate first-principles methods properly capturing excitonic effects for both spectroscopies are necessary.

In a static formulation, the oscillator strength of a phonon mode in the IR spectrum can be computed from the Born effective charge that is defined as the derivative of macroscopic polarization with respect to this mode. Raman scattering intensity, on the other hand, can be expressed as derivatives of polarizability tensors with respect to phonon modes within the Placzek approximation [14]. Both approaches have been implemented in first-principles codes [15, 16]. The theory and calculations of the IR spectrum was extended to metals with an emphasis on non-adiabatic effects recently [17]. *Ab initio* calculations of Raman scattering intensity have also been routinely performed for various materials, although its applicability to resonant excitations (crucial to study laser frequency dependence) was debated [18]. Development of Raman scattering theory including excitonic effects is still in its infancy. Based on a finite-displacement method, first-principles calculations were conducted for both first-order [8, 11, 19] and second-order resonant Raman scattering [20]. A new diagrammatic approach including non-adiabatic effects was reported and applied to bulk h-BN and monolayer $MoS_2$ [21]. However, to the best of our knowledge, excitonic effects on IR spectrum have not been discussed from first principles.

In this work, we derived an expression for both the IR spectrum and Raman scattering cross-section including excitonic effects by means of a perturbation theory in the framework of the time-dependent adiabatic-GW (TD-aGW) theory [22, 23]. For the IR spectrum, our expression reduces to the results from a time-dependent formulation of density functional perturbation theory (DFPT) [24] in the limiting case when excitonic effects are omitted. Exciton effects in fact dress the electron-hole propagators as we demonstrate below. Our derived expression with excitonic effects included can be seen as replacing the free electron-hole propagators with the interacting ones. For Raman scattering intensity, our results partially agree with those in Ref. [21] but include an additional term responsible for dressing the electron-phonon vertex by excitonic effects.

The rest of the paper is organized as following. In Sec. II, we introduce the Hamiltonian with electron-phonon and electron-light couplings in the single-particle orbital basis. The equation of motion (EOM) of the single-electron density matrix and phonon displacement without excitonic effects are given in Sec. II A, from which we solved for the IR spectrum and Raman scattering intensity spectrum without electron-hole interactions. In Sec. II B, we introduce the EOM with excitonic effects included within the TD-aGW theory. Expressions for both IR spectrum and Raman intensity are derived with an interacting Green function approach. We apply our method to monolayer $MoS_2$, $WS_2$, and $WSe_2$ in Sec. III and show that the exciton- phonon coupling and exciton energy landscape lead to very distinct features in the Raman spectra of these three structurally similar materials. We discuss higher-order electron-phonon coupling and some effects beyond perturbation theory and summarize our work in Sec. IV.

## II.   THEORY

We start with the non-interacting Hamiltonian of the combined electron and phonon system,

$$H_0 = \sum_{nk} \epsilon_{nk} a_{nk}^\dagger a_{nk} + \sum_{q\nu} \hbar \omega_{q\nu} \left( b_{q\nu}^\dagger b_{q\nu} + \frac{1}{2} \right)$$

$$(1)$$

, where $a_{nk}^\dagger$ ($a_{nk}$) is the creation (annihilation) operator of the $n$-th band electron with crystal momentum $\boldsymbol{k}$, $b_{q\nu}^\dagger$ is a phonon creation operator of the mode $\nu$ and wavevector $\boldsymbol{q}$, and $\epsilon_{nk}$ and $\omega_{q\nu}$ are the corresponding electron and phonon energy, respectively. Both electron and phonon energies can be obtained at different levels of theory, e.g., from standard density functional theory (DFT) or GW quasiparticle calculations. Electron-phonon (e-ph) couplings are included as a first order change in the electron energy with respect to atomic displacements, with

$$H_{e-ph} = \frac{1}{\sqrt{N_p}} \sum_{nmk q\nu} g_{nm\nu}(\boldsymbol{k}, \boldsymbol{q}) \left( b_{q\nu} + b_{-q\nu}^\dagger \right) a_{nk+q}^\dagger a_{mk},$$

$$(2)$$

where $N_p$ is the size of the Born-von Karman supercell and $g_{nm\nu}(\boldsymbol{k}, \boldsymbol{q})$ is the e-ph coupling matrix element between a electronic state of a band index n, momentum $\boldsymbol{k} + \boldsymbol{q}$ and another state with a band index m, momentum $\boldsymbol{k}$, which are coupled via $\nu$-th phonon mode with wavevector q. E-ph coupling matrix elements can be computed from DFPT [25] or the advanced GWPT [26, 27] with GW self-energy corrections from first-principles [5,7]. For the first order IR spectrum and Raman scattering cross-section, we consider only zone center ($\boldsymbol{q} = 0$) phonons.

Electron-light (e-l) couplings are treated semi-classically. We write this coupling term in the length gauge [28] and in the long wavelength limit with the dipole approximation,

$$H_L(t) = e\boldsymbol{E}(t) \cdot \sum_{ijk} \boldsymbol{d}_{ijk} a_{ik}^\dagger a_{jk},$$

$$(3)$$

where e is the magnitude (positive value) of the electron charge, $\boldsymbol{E}(t)$ is a time-dependent external field, and $\boldsymbol{d}_{ijk}$ is the optical matrix element between single-particle Bloch states with band indices $i, j$ at the momentum $\boldsymbol{k}$.

## A.   Free electron-hole pairs

In the case of neglecting excitonic effects, the expressions for calculating of IR spectrum had been derived from time-dependent density functional perturbation theory (TD-DFPT) [17, 24] or a diagrammatic approach [29, 30], while Raman scattering cross-section including nonadiabatic effects had also been derived diagrammatically [31–33]. Our derivation is based on a time-dependent perturbation theory in the one-particle density matrix formulation. We start with a classical treatment to the phonon displacement in Eq. (2). The equation of motion (EOM) of the single-electron density matrix $\rho$ in the Bloch state basis without electron-hole (e-h) interactions reads (note that $\rho$ is diagonal in $\boldsymbol{k}$ because of the assumptions of $\mathbf{q}=0$ phonons and long-wavelength light field),

$$i\hbar \frac{\partial}{\partial t} \rho_{nm\boldsymbol{k}}(t) = [h^{cph}(t), \rho(t)]_{nm\boldsymbol{k}},$$

(4)

where the time-dependent effective Hamiltonian $h^{cph}(t)$, after taking a classical approximation to the displacement operator $Q_\nu(t) = (b_{0\nu} + b_{0\nu}^\dagger)/\sqrt{2}$ in the e-ph coupling term in Eq. (2), is given by

$$h_{nm\boldsymbol{k}}^{cph}(t) = \epsilon_{n\boldsymbol{k}} \delta_{nm} + \left( \sum_\nu \tilde{g}_{nm\nu}(\boldsymbol{k}, \boldsymbol{q}=0) \langle Q_\nu(t) \rangle \right) + e\boldsymbol{E}(t) \cdot \boldsymbol{d}_{nm\boldsymbol{k}},$$

(5)

where $\langle Q_\nu(t) \rangle$ is the time-dependent expectation value of the displacement operator. To keep the notation simpler, we further define $\tilde{g}_{nm\nu\boldsymbol{k}} = \sqrt{2} N_p^{-\frac{1}{2}} g_{nm\nu}(\boldsymbol{k}, \boldsymbol{q}=0)$. For the phonon part, the EOM of $\langle Q_\nu(t) \rangle$ reads,

$$\left( \frac{\partial^2}{\partial t^2} + \omega_{0\nu}^2 \right) \langle Q_\nu(t) \rangle = -\frac{\omega_{0\nu}}{\hbar} \sum_{nm\boldsymbol{k}} \tilde{g}_{nm\nu\boldsymbol{k}} \rho_{mn\boldsymbol{k}}(t).$$

(6)

The detailed derivations of Eq. (4) and (6) are given in the Appendix 1. The formal solution of $\langle Q_\nu(\omega) \rangle$ in the frequency domain can be written down as

$$\langle Q_\nu(\omega) \rangle = \frac{-\omega_{0\nu}}{\hbar(-\omega^2 + \omega_{0\nu}^2)} \sum_{nm\boldsymbol{k}} \tilde{g}_{nm\nu\boldsymbol{k}} \rho_{mn\boldsymbol{k}}(\omega) = D_{0\nu}(\omega) \sum_{nm\boldsymbol{k}} \tilde{g}_{nm\nu\boldsymbol{k}} \rho_{mn\boldsymbol{k}}(\omega),$$

(7)

where we make use of standard notation for the bare phonon propagator of the mode $\nu$, $D_{0\nu}(\omega)$ [34]. Eq. (4)-(6) are coupled equations of the time evolution of the single electron density matrix and phonon displacement expectation values. We will solve the dynamics by treating e-l and e-ph couplings as perturbations.

### 1. IR spectrum in the independent-electron framework

We write Eq. (4) in the frequency domain and arrange the coupling terms to the right-hand side,

$$(\hbar\omega - \epsilon_{nm\boldsymbol{k}}) \rho_{nm\boldsymbol{k}}(\omega) = \int \frac{d\Omega}{2\pi} \left[ \sum_\nu \langle Q_\nu(\omega - \Omega) \rangle \, \tilde{g}_\nu + e\boldsymbol{E}(\omega - \Omega) \cdot \boldsymbol{d}, \rho(\Omega) \right]_{nm\boldsymbol{k}}.$$

(8)

where we define $\epsilon_{nm\boldsymbol{k}} = \epsilon_{n\boldsymbol{k}} - \epsilon_{m\boldsymbol{k}}$. In perturbation theory, solution for coupling to the (n+1)-th order is solved by inserting the n-th order solution on the right-hand side of Eq. (8). We consider a semiconductor at low temperature, in which case the zeroth order solution is $\rho_{vv\boldsymbol{k}}^{(0)} = 1$ for all valence bands $v$ and all other components, $\rho_{cc'\boldsymbol{k}}^{(0)}$, $\rho_{vv'\boldsymbol{k}}^{(0)}$, and $\rho_{cvk}^{(0)}$ are zero. (Here $c$ stands for the conduction bands.) To first order of the external light field alone, the density matrix solution for a general pair of $n$, $m$ bands is

$$\rho_{nmk}^{(l)}(\omega) = \frac{e\boldsymbol{E}(\omega) \cdot \boldsymbol{d}_{nmk} f_{mnk}}{\hbar\omega - \epsilon_{nmk}},$$

<div align="right">(9)</div>

where $f_{mnk} = f_{mk} - f_{nk}$ with f the Fermi-Dirac distribution function. The superscript l (which stands for interaction with light) indicates that the solution is to the first order of the e-l coupling. The off-diagonal matrix elements (in the band indices) of Eq. (9) describe the coherence induced by the field, from which we can compute the electron polarization and obtain the dielectric function [22, 28].

For the IR absorption spectrum, we are interested in how the ion displacements and e-ph interactions induce changes in coherence and thus the polarization. The lowest order coherence solution with a mix of e-l and e-ph couplings can be obtained from the first term in the bracket on the right-hand side (r.h.s.) of Eq. (8). For the lowest order $\langle Q_\nu \rangle$ solution driven by the external field, we insert Eq. (9) to the r.h.s. of Eq. (6) and write

$$\langle Q_\nu(\omega) \rangle = eD_{0\nu}(\omega) \sum_{ilk'} \frac{\tilde{g}_{li\nu k'}\boldsymbol{E}(\omega) \cdot \boldsymbol{d}_{ilk'} f_{lik'}}{\hbar\omega - \epsilon_{ilk'}}.$$

<div align="right">(10)</div>

Inserting Eq. (10) and setting $\rho(\Omega)$ to its zeroth order solution in Eq. (8), we get the solution to first order in both the e-l and e-ph couplings,

$$\rho_{nmk}^{(Ql)}(\omega) = e\sum_{\nu ilk'} [\frac{\tilde{g}_{nm\nu k} f_{mnk}}{\hbar\omega - \epsilon_{nmk}} D_{0\nu}(\omega) \frac{\tilde{g}_{li\nu k'}\boldsymbol{E}(\omega) \cdot \boldsymbol{d}_{ilk'} f_{lik'}}{\hbar\omega - \epsilon_{ilk'}}],$$

where the superscript $(Q, l)$ indicates the solution is obtained by an ordered sequence of e-l and e-ph perturbations. We note that $\rho_{cc'}$ and $\rho_{\nu\nu'}$ vanish at this order.

From the perturbed density matrix we obtain the induced polarization change,

$$\Delta P^\mu(\omega) = -\frac{e}{N_pV} \sum_{nmk} d_{nmk}^\mu \rho_{mnk}^{(Ql)}(\omega) = -\frac{e^2}{N_pV} \sum_{nmk\nu} [\frac{\tilde{g}_{nm\nu k} d_{mnk}^\mu f_{mnk}}{\hbar\omega - \epsilon_{nmk}} D_{0\nu}(\omega) \sum_{\alpha ilk'} \frac{\tilde{g}_{li\nu k'} E^\alpha(\omega) d_{ilk'}^\alpha f_{lik'}}{\hbar\omega - \epsilon_{ilk'}}],$$

where $\mu, \alpha$ are indices of the Cartesian coordinates and $V$ is the unit cell volume. The corresponding susceptibility is computed from $\chi^{\mu\alpha}(\omega) = \frac{P^\mu(\omega)}{\epsilon_0 E^\alpha(\omega)}$ with

$$\chi^{\mu\alpha}(\omega) = \frac{e^2}{\epsilon_0 N_k V} \sum_\nu \frac{\not{p}_\nu^\mu(\omega) \not{p}_\nu^\alpha(\omega)}{\omega_{0\nu}^2 - \omega^2},$$

<div align="right">(11)</div>

where $\epsilon_0$ is the vacuum permittivity and we define the IR oscillator strength $\not{p}_\nu^\mu$ as

$$\not{p}_\nu^\mu(\omega) = \sqrt{\frac{\omega_{0\nu}}{\hbar}} \sum_{ilk'} \frac{\tilde{g}_{li\nu k'} d_{ilk'}^\mu f_{lik'}}{\hbar\omega - \epsilon_{ilk'}}.$$

<div align="right">(12)</div>

Eq. (11) agrees with the derivation from TD-DFPT in Ref. [17, 24], where the oscillator strength is expressed in terms of the dynamical Born effective charge. To see the equivalence, we can rewrite the dynamical Born effective charge in terms of e-ph coupling and optical matrix elements. The details are given in the Appendix 2.

The physical mechanism of the electronic part of the IR response can be clearly seen in the derivation. It is the response of the dynamical polarization due to the perturbed lattice vibration; hence it is tightly related to the dynamical Born effective charge discussed in Ref. [24, 35]. In contrast to the derivation from a static perturbation theory, non-adiabatic effects are naturally included in our time-dependent theory. A direct correspondence to the diagrams shown in Ref. [29, 30] can also be identified. The first bubble represents an e-h pair generated by the external field, which later recombines and induces a phonon displacement. The second bubble stands for an e-h pair generated by the phonon displacement through e-ph couplings then recombines to emit light.

## 2. Raman scattering in the independent-electron framework

Raman scattering cross-section can be derived by assuming that $\langle Q_v(t)\rangle$ admits a free propagator solution or coherent phonon motion. The scattered light generated from the polarization field is modulated by coherent phonon motion through e-ph couplings. At the lowest order, we focus on solutions from one-photon and one-phonon perturbations. Generalization to higher order Raman scatterings is possible but involves density matrix elements of states differed by finite crystal momentum (i.e., $\boldsymbol{k} \neq \boldsymbol{k}'$).

With the assumption of a coherent phonon motion, the role of $\langle Q_v(t)\rangle$ is equivalent to an independent external field. In second-order perturbation theory, we shall include all possible combinations of e-l and e-ph couplings. We consider the response where the system is first excited by a light field and then modulated by coherent phonons. Substituting $\langle Q_v\rangle$ and $\rho_{nm}$ in the first term on the right-hand side of Eq. (8) with the free propagator solution $\langle Q_v(\omega) = Q_{0v}(\delta(\omega - \omega_{0v}) + \delta(\omega + \omega_{0v}))$ and Eq. (9), respectively, we have

$$\rho_{nm\boldsymbol{k}}^{(Q_c l)}(\omega) = \sum_{\alpha=+,\mu\nu} \frac{eQ_{0v}E^\mu(\omega+\alpha\omega_{0v})}{\hbar\omega - \epsilon_{nm\boldsymbol{k}}} \sum_j \left[\frac{\tilde{g}_{njv\boldsymbol{k}}d_{jm\boldsymbol{k}}^\mu f_{mj\boldsymbol{k}}}{\hbar(\omega+\alpha\omega_{0v})-\epsilon_{jm\boldsymbol{k}}} - \frac{d_{nj\boldsymbol{k}}^\mu \tilde{g}_{jmv\boldsymbol{k}}f_{jn\boldsymbol{k}}}{\hbar(\omega+\alpha\omega_{0v})-\epsilon_{nj\boldsymbol{k}}}\right],$$

where the $(Q_c l)$ in the superscript indicates the order of perturbations and $Q_c$ denotes the coherent phonon perturbation. The $\boldsymbol{E} \cdot \boldsymbol{d}$ term on the right-hand side of Eq. (8) generates second order optical responses, such as second harmonic generation, which has been extensively discussed in the literature [28] but does not contribute to Raman scattering at this order.

In the case where the system is initially driven by coherent phonons, to the first order in the e-ph coupling, we have

$$\rho_{nm\boldsymbol{k}}^{(Q_c)} = \sum_v \frac{\langle Q_v(\omega)\rangle f_{mn\boldsymbol{k}}\tilde{g}_{nmv\boldsymbol{k}}}{\hbar\omega - \epsilon_{nm\boldsymbol{k}}}.$$

$(13)$

A subsequent coupling to the external field in turn generates the next order coherence. By inserting $\rho_{nm\boldsymbol{k}}^{(Q_c)}(\omega)$ to the $\boldsymbol{E} \cdot \boldsymbol{d}$ term in the r.h.s of Eq. (8), we obtain

$$\rho_{nm\boldsymbol{k}}^{(lQ_c)}(\omega) = e \sum_{\alpha=+,\mu\nu} \frac{Q_{0v}E^\mu(\omega+\alpha\omega_{0v})}{\hbar\omega - \epsilon_{nm\boldsymbol{k}}} \sum_j \left[-\frac{f_{jn\boldsymbol{k}}\tilde{g}_{njv\boldsymbol{k}}d_{jm\boldsymbol{k}}^\mu}{-\alpha\hbar\omega_{0v}-\epsilon_{nj\boldsymbol{k}}} + \frac{f_{mj\boldsymbol{k}}\tilde{g}_{jmv\boldsymbol{k}}d_{nj\boldsymbol{k}}^\mu}{-\alpha\hbar\omega_{0v}-\epsilon_{jm\boldsymbol{k}}}\right].$$

The second-order phonon-modulated density matrix solution is obtained by combining the two contributions,

$$\rho_{nm\boldsymbol{k}}^{(2)}(\omega) = \rho_{nm\boldsymbol{k}}^{(Q_c l)}(\omega) + \rho_{nm\boldsymbol{k}}^{(lQ_c)}(\omega).$$

$(14)$

The Raman scattering cross-section can be computed by considering the light radiated from the polarization oscillations following Loudon [32]. The radiated field at position $\boldsymbol{r}$ generated by polarization oscillations at the origin is

$$E^\mu(\boldsymbol{r}, t) = -\frac{\omega_D^2}{4\pi\epsilon_0 c^2 |\boldsymbol{r}|} P^\mu\left(t - \frac{|\boldsymbol{r}|}{c}\right).$$

The differential cross-section of the scattered light at a detected frequency $\omega_D$ from an incident light of frequency $\omega_{in}$ is

$$\frac{d\sigma}{d\Omega} = \frac{r^2 I_D \omega_{in}}{I_{in}\omega_D},$$

where $I_D$ and $I_{in}$ are the intensity of the scattered light and the incident light field, respectively. The energy fraction accounts for the fact that part of the light energy is removed from the incident beam. We thus have

$$\frac{d\sigma}{d\Omega}(\omega_D, \omega_{in}) = \frac{\frac{\epsilon_0 c}{2} \frac{\omega_D^4}{16\pi^2 \epsilon_0^2 c^4} \frac{\omega_{in}}{\omega_D} |P(\omega_D)|^2}{\frac{\epsilon_0 c}{2} |E(\omega_{in})|^2} = \frac{\omega_D^3 \omega_{in}}{16\pi^2 \epsilon_0^2 c^4} |M(\omega_D, \omega_{in})|^2,$$

(15)

where we define the scattering matrix element, $M_{\alpha,\mu}(\omega_D, \omega_{in})$, which can be compared to that in Eq. (1) in Ref. [36].

For the Stoke process we evaluate the change in the polarization at $\omega_D = \omega_{in} - \omega_{0\nu}$ in Eq. (14). The final expression of the scattering matrix element is

$$M_{\alpha,\mu}(\omega_D, \omega_{in}) = \sum_{\nu} -\frac{e^2 Q_{0\nu}}{N_p V} \{ \sum_{nmjk} [\frac{d_{mnk}^{\alpha} \tilde{g}_{njvk} d_{jmk}^{\mu}}{\hbar\omega_D - \epsilon_{nmk}} \left( \frac{f_{mjk}}{\hbar\omega_{in} - \epsilon_{jmk}} - \frac{f_{jnk}}{-\hbar\omega_{0\nu} - \epsilon_{njk}} \right)$$
$$+ \frac{d_{mnk}^{\alpha} \tilde{g}_{jmvk} d_{njk}^{\mu}}{\hbar\omega_D - \epsilon_{nmk}} \left( -\frac{f_{jnk}}{\hbar\omega_{in} - \epsilon_{njk}} + \frac{f_{mjk}}{-\hbar\omega_{0\nu} - \epsilon_{jmk}} \right) ]\}$$

(16)

Our result agrees with the first-order Raman cross-section within the independent electron picture given in Ref. [33] and Ref. [37]. We note that two out of the six terms in Ref. [37] are beyond the semi-classical treatment of e-l couplings used here. Nevertheless, the dominant double resonance terms which are the first and the third terms in the square bracket in Eq. (16), are captured in our formulation.

## B. Excitonic effects

To study excitonic effects on both IR and Raman spectrum, we include a GW self-energy term in the time-dependent Hamiltonian as in real-time propagation TD-aGW theory [22, 23]. In the adiabatic approximation, the time-dependent change in the GW self-energy is replaced by that of a static COHSEX self-energy. Accordingly, equilibrium GW quasi-particle energy is used as the unperturbed Hamiltonian for Eq. (1). The e-ph matrix elements are taken at the DFPT level in this work. The many-electron interaction term in the TD-aGW theory reads

$$H_{ee}(t) = \delta V^H(t) + \delta \Sigma^{COHSEX}(t),$$

where $\delta V^H$ is the time-dependent change in the Hartree potential from the equilibrium and $\Sigma^{COHSEX}$ is the change in the static COHSEX self-energy due to the driving fields which is given by the screened Coulomb interaction. Both terms are functionals of the density matrix. We can write the matrix elements of $H_{ee}$ in the Bloch state basis as

$$[H_{ee}(t)]_{ijk} = \sum_{nmk'} K_{ijk,nmk'} \rho_{nmk'}(t),$$

where $K_{ijk,nmk'}$ is the kernel matrix elements including the bare and the screened Coulomb potential derived from Hartree and static COHSEX self-energy, respectively. Explicit expressions of the self-energy and kernel matrix elements can be found in Ref. [22] and [23]. The kernel K aside of a minus sign in fact is the kernel that goes into the standard GW-BSE formulation for calculations of excitonic states and linear optical response of solids.[6] It has been shown that the first-order response function for optical absorption without e-ph coupling terms is equivalent to that computed from the state-of-art GW-BSE calculations [22]. Thus, having the term $H_{ee}(t)$ in TD-aGW includes excitonic effects in the theory.

With in the Tamm-Dancoff approximation, the EOM of the elements of the density matrix $\rho_{cvk}$ is given by [38]

$$i\hbar \frac{\partial \rho_{cvk}(t)}{\partial t} - \sum_{c'v'k'} H_{cvk,c'v'k'} \rho_{c'v'k'}(t) = \sum_{\nu} [\tilde{g}_{\nu} \langle Q_{\nu}(t) \rangle + e\boldsymbol{E} \cdot \boldsymbol{d}, \rho(t)]_{cvk},$$

(17)

where $H_{cvk,c'v'k'}$ is the effective two-particle Hamiltonian of the Bethe-Salpeter equation (BSE),

$$H_{cvk,c'v'k'} = \epsilon^{QP}_{cvk}\delta_{cc'}\delta_{vv'}\delta_{kk'} - K_{cvk,c'v'k'},$$

(18)

with $\epsilon^{QP}_{cvk}$ the difference in GW quasi-particle b a n d energy in the conduction ($c$) and valence band ($v$) states at wavevector $k$. In writing down the effective Hamiltonian on the r.h.s of Eq. (17), we use the property that the first order occupation change is negligible so that only the zeroth order $\rho_{vvk}$ is nonzero.

In Ref. [38], Eq. (17) was solved perturbatively for the case *without* e-ph couplings. Following the same procedure, we introduce the Green's function,

$$\mathcal{G}_{cvk,c'v'k'}(\omega) = \sum_s \frac{A^{(s)}_{cvk}A^{(s)*}_{c'v'k'}}{\hbar\omega - E_s},$$

where $E_s$ and $A^{(s)}_{cvk}$ are the excitation energy and the **k**-space envelope function of the $s$-th exciton state, respectively, from solving the BSE with the Hamiltonian given by Eq. (18). With the help of the Green's function, an approach similar to that given above in Sec. II A can be applied for the perturbed solutions of the density matrix in Eq. (17) in the presence of the e-ph coupling terms. Explicitly, the formal solution of Eq. (17) in the frequency domain reads

$$\rho_{cvk}(\omega) = \sum_{c'v'k'} \mathcal{G}_{cvk,c'v'k'}(\omega)\int\frac{d\Omega}{2\pi}\left[\sum_{mv}\left(\tilde{g}_{c'mvk'}\rho_{mv'k'}(\omega - \Omega) - \tilde{g}_{mv'vk'}\rho_{c'mk'}(\omega - \Omega)\right)\langle Q_v(\Omega)\rangle\right.$$
$$\left.+ e\sum_{m\mu}E^\mu(\Omega)\left(\rho_{mv'k'}(\omega - \Omega)d^\mu_{c'mk'} - \rho_{c'mk'}(\omega - \Omega)d^\mu_{mv'k'}\right)\right].$$

(19)

Here m is an index running over all bands. To the lowest order of e-l and e-ph couplings, we obtain

$$\rho^{(1)}_{cvk}(\omega) = \rho^{(Q)}_{cvk}(\omega) + \rho^{(l)}_{cvk}(\omega) = \sum_{c'v'k'}\mathcal{G}_{cvk,c'v'k'}(\omega)\left[\sum_v\tilde{g}_{c'v'vk'}\langle Q_v(\omega)\rangle + \sum_\mu eE^\mu(\omega)d^\mu_{c'v'k'}\right],$$

(20)

where the superscript $l$ and $Q$ indicate the solution is from e-l and e-ph couplings, respectively, and we have used $\rho^{(0)}_{cc'}(\omega) = \rho^{(0)}_{cv}(\omega) = 0$ and $\rho^{(0)}_{vv'}(\omega) = 2\pi\delta_{vv'}\delta(\omega)$ for a semiconductor at low temperature. An expression for the linear absorption spectrum including excitonic effects can be derived from the second term in the bracket in Eq. (20) only.

### 1. IR spectrum including excitonic effects

For the IR spectrum, we again solve for the light-induced displacement and the change in polarization induced by it through e-ph couplings. The EOM of the classical displacement $\langle Q_v(t)\rangle$, Eq. (6), and its formal solution, Eq. (7), do not change upon including the added term $H_{ee}(t)$ (i.e., excitonic effects). Inserting the second term in the r.h.s. of Eq. (20) to Eq. (7), the lowest order displacement solution driven by an optical field is

$$\langle Q_\lambda(\omega)\rangle = D_{0\lambda}(\omega)\sum_{cvk}[\tilde{g}_{cv\lambda k}\rho^{(l)}_{vck}(\omega) + \tilde{g}_{vc\lambda k}\rho^{(l)}_{cvk}(\omega)]$$
$$= D_{0\lambda}(\omega)\sum_{cvk}[(\tilde{g}_{cv\lambda k}\sum_{c'v'k'}\mathcal{G}_{cvk,c'v'k'}(\omega)\sum_\mu eE^\mu(\omega)d^\mu_{c'v'k'}) + (\omega \leftrightarrow -\omega)^*].$$

(21)

In the last line, we use the property that matrix elements $\rho_{vck}$ can be obtained from the relation $\rho_{vck}(\omega) = \rho_{cvk}(-\omega)^*$ and the notation $(\omega \leftrightarrow -\omega)^*$ means a similar expression as the preceding term but it is its complex conjugate with $\omega$ changed to $-\omega$. Following the general procedure in Sec. II A, the lowest order coherence modulated by e-ph couplings is obtained by substituting Eq. (21) into the first term in the bracket of Eq. (20)

$$\rho_{cv\mathbf{k}}^{(QI,ee)} = e \sum_{\mu\nu} E^\mu(\omega) D_{0\lambda}(\omega) \sum_s \frac{A_{cv\mathbf{k}}^{(s)} G_{s\lambda}}{\hbar\omega - E_s} \sum_{s'} \left[ \frac{R_{s'}^\mu G_{s'\lambda}^*}{\hbar\omega - E_{s'}} + (\omega \leftrightarrow -\omega)^* \right],$$

where we define excitonic version of matrix elements, $R_s^\mu \equiv \sum_{c'v'\mathbf{k}'} A_{c'v'\mathbf{k}'}^{(s)*} d_{c'v'\mathbf{k}'}^\mu$ and $G_{s'\lambda}^* \equiv \sum_{c_1v_1\mathbf{k}_1} \tilde{g}_{v_1c_1\lambda\mathbf{k}_1} A_{c_1v_1\mathbf{k}_1}^{(s')}$, both of which can be viewed as matrix elements of a coupling term between the many-electron ground state and exciton states.

The susceptibility change induced at this order reads,

$$\chi^{\mu\alpha}(\omega) = \frac{e^2}{\epsilon_0 N_p V} \sum_\nu \frac{F_\nu^\mu(\omega) F_\nu^{\alpha*}(\omega)}{\omega_{0\nu}^2 - \omega^2},$$

$(22)$

where we define the excitonic oscillator strength for the IR spectrum,

$$F_\nu^\mu(\omega) = \sqrt{\frac{\omega_{0\nu}}{\hbar}} \sum_s \frac{G_{s\nu}^* R_s^\mu}{\hbar\omega - E_s} + (\omega \leftrightarrow -\omega)^*.$$

$(23)$

Diagrammatically, this result can be interpreted similarly as its counterpart in Eq. (11). With e-h interactions, free e-h propagators in Eq. (11) are replaced by interacting e-h propagators from solutions of the BSE. In the limit of the non-interacting case where exciton envelope functions are replaced by a $\mathbf{k}$-space delta function and exciton energy are replaced by free e-h energy, we recover the results in Sec. II A.

### 2. Raman scattering cross-section including excitonic effects

The derivation for Raman scattering cross-section including excitonic effects can be done following the general procedure in Sec. II A 2. We set the solution of the phonon displacement $\langle Q_\nu(\omega) \rangle$ as a free harmonic oscillator and derive the polarization modulated by the coherent phonon motions. The lowest nontrivial coherence term induced by excitations from the light field and later perturbed by electron-phonon couplings reads,

$$\begin{aligned}
\rho_{cv\mathbf{k}}^{(Q_cI,eh)}(\omega) &= \sum_{\nu c'v'\mathbf{k}'} \mathcal{G}_{cv\mathbf{k},c'v'\mathbf{k}'}(\omega) \int \frac{d\Omega}{2\pi} \left( \sum_{c_1} \tilde{g}_{c'c_1\nu\mathbf{k}'} \rho_{c_1v'\mathbf{k}'}^{(l)}(\Omega) - \sum_{v_1} \tilde{g}_{v_1v'\nu\mathbf{k}'} \rho_{c'v_1\mathbf{k}'}^{(l)}(\Omega) \right) \langle Q_\nu(\omega - \Omega) \rangle \\
&= e \sum_{s\nu c'v'\mathbf{k}'} \frac{A_{cv\mathbf{k}}^{(s)} A_{c'v'\mathbf{k}'}^{(s)*}}{\hbar\omega - E_s} \sum_{\mu c''v''\mathbf{k}''} \int d\Omega \left( \sum_{c_1} \tilde{g}_{c'c_1\nu\mathbf{k}'} \mathcal{G}_{c_1v'\mathbf{k}',c''v''\mathbf{k}''}(\Omega) - \sum_{v_1} \tilde{g}_{v_1v'\nu\mathbf{k}'} \mathcal{G}_{c'v_1\mathbf{k}',c''v''\mathbf{k}''}(\Omega) \right) \langle Q_\nu(\omega \\
&\quad - \Omega) \rangle d_{c''v''\mathbf{k}''}^\mu E^\mu(\Omega) = e \sum_{ss'\nu\mu} \frac{A_{cv\mathbf{k}}^{(s)} G_{ss'\nu} R_{s'}^\mu}{\hbar\omega - E_s} \sum_{\alpha=\pm} \frac{Q_{0\nu} E^\mu(\omega + \alpha\omega_{0\nu})}{\hbar(\omega + \alpha\omega_{0\nu}) - E_{s'}},
\end{aligned}$$

where we insert the first order solution perturbed by external light field in the second line and define the exciton-phonon coupling matrix elements,

$$G_{ss'\nu} = \sum_{c'v'\mathbf{k}'} A_{c'v'\mathbf{k}'}^{(s)*} \left( \sum_{c_1} \tilde{g}_{c'c_1\nu\mathbf{k}'} A_{c_1v'\mathbf{k}'}^{(s')} - \sum_{v_1} \tilde{g}_{v_1v'\nu\mathbf{k}'} A_{c'v_1\mathbf{k}'}^{(s')} \right),$$

$(24)$

in the last line [39, 40].

At the same order of the perturbation theory, the response can also be induced first by the coherent phonon motion, where we have,

$$\rho_{cv\mathbf{k}}^{(Q_c,eh)}(\omega) = \sum_{c'v'\mathbf{k}'\nu} \mathcal{G}_{cv\mathbf{k},c'v'\mathbf{k}'}(\omega)\tilde{g}_{c'v\nu\mathbf{k}'}\langle Q_\nu(\omega)\rangle.$$

In the next order, the induced coherence is further coupled to the external light field, and we obtain,

$$\rho_{cv\mathbf{k}}^{(lQ_c,eh)}(\omega) = e\sum_{ss'\mu\nu}\frac{A_{cv\mathbf{k}}^{(s)}X_{ss'}^\mu G_{s'\nu}}{\hbar\omega - E_s}\sum_{\alpha=\pm}\frac{Q_{0\nu}E^\mu(\omega+\alpha\omega_{0\nu})}{-\alpha\hbar\omega_{0\nu}-E_{s'}},$$

where in the last line we define the exciton-dipole coupling matrix elements,

$$X_{ss'}^\mu = \sum_{c'v'\mathbf{k}'}A_{c'v'\mathbf{k}'}^{(s)*}\left(\sum_{j\in c}d_{c'j\mathbf{k}'}^\mu A_{jv'\mathbf{k}'}^{(s')} - \sum_{i\in v}d_{iv'\mathbf{k}'}^\mu A_{c'i\mathbf{k}'}^{(s')}\right),$$

and use $G_{s'\nu}^* = \sum_{v_1c_1\mathbf{k}_1}\tilde{g}_{v_1c_1\nu\mathbf{k}_1}A_{c_1v_1\mathbf{k}_1}^{(s')}$. The physical meaning of $X_{ss'}^\mu$ can be understood by drawing an analogy from the exciton-phonon coupling matrix elements, Eq. (24). The first term in the bracket describes electrons in two excitons coupled via the field of the light, while the second term describes the same coupling for holes in two excitons.

Finally, combining the two contributions, we write down the phonon mode resolved Raman scattering matrix elements for the Stoke process,

$$M_{\alpha\mu}^\nu(\omega_D,\omega_{in}) \propto Q_{0\nu}\Big[\sum_{ss'}\frac{R_s^{\alpha*}\mathcal{G}_{ss'\nu}R_{s'}^\mu}{(\hbar\omega_D - E_s)(\hbar(\omega_D+\omega_{0\nu})-E_{s'})} + \sum_{ss'}\frac{R_s^\alpha \mathcal{G}_{ss'\nu}R_{s'}^{\mu*}}{(-\hbar\omega_D - E_s)(\hbar(-\omega_D-\omega_{0\nu})-E_{s'})}$$
$$+ \sum_{ss'}\frac{R_s^{\alpha*}X_{ss'}^\mu G_{s'\nu}}{(\hbar\omega_D - E_s)(-\hbar\omega_{0\nu}-E_{s'})} + \sum_{ss'}\frac{R_s^\alpha X_{ss'}^\mu G_{s'\nu}^*}{(-\hbar\omega_D - E_s)(\hbar\omega_{0\nu}-E_{s'})}\Big].$$

$$(25)$$

The first two terms agree with the results derived from a diagrammatic method in Ref. [21]. For resonant excitations, the first two terms are the double resonance terms since phonon energy is small compared to the exciton excitation energy [33]. Diagrammatically, these two terms correspond to dressing the photon vertex with ladder diagrams but keeping the e-ph vertex as those computed from DFPT. The additional two terms in our derivation correspond to a choice of dressing one e-ph and one photon vertex with ladder diagrams. We note importantly that the r.h.s. of Eq. (25) is directly proportional to the coherent phonon motion amplitude $Q_{0\nu}$, which depend on external conditions such as temperature and excitations.

## III. APPLICATIONS

To demonstrate the excitonic effects on IR and resonant Raman spectrum, we apply our method to monolayer $MoS_2$, $WS_2$, and $WSe_2$. We implemented Eq. (11) and Eq. (22) for the IR spectrum with and without excitonic effects, respectively, and Eq. (16) and Eq. (25) for the Raman spectrum with and without excitonic effects, respectively, into computer codes. The ingredients for the calculations are obtained with open-source packages -- Quantum Espresso [41] for DFT quantities, EPW [42] for electron-phonon couplings, and BerkeleyGW [5–7] for GW quasiparticle energies and exciton energies and wavefunctions.

We performed DFT calculations with PBE pseudopotentials [43, 44] from the SG15 ONCV potentials database [45]. For the ground-state calculation, we use a **k**-mesh of 12×12×1 and a plane wave energy cutoff of 50 Ry. A vacuum of 20 Å is chosen to prevent spurious interactions between periodic images. The GW quasiparticle energies, exciton energies and wavefunctions (k-space envelope functions) are computed with the BerkeleyGW package. A **k**-grid of 24×24×1 with a subsampling of 10 points in the mini-Brillouin zone [46] and a dielectric energy cutoff of 10 Ry and 5000 bands are used in the GW calculation for $MoS_2$. For $WS_2$ and $WSe_2$, we use a **k**-grid of 12×12×1 with a subsampling of 10 points. The dielectric energy cutoff of 10 Ry and band cutoff of 2000 bands are used. The frequency-dependence of the dielectric screening is computed using the Hybertsen-Louie generalized plasmon pole model [5]. The Bethe-Salpeter equations (BSE) are solved, with the electron-hole interaction kernel from the GW calculations, on a uniform 24×24 **k**-grid with 8 conduction and 8 valence bands for $MoS_2$ and on a 36×36 **k**-grid with 6 conduction and 6 valence bands for both $WS_2$ and $WSe_2$. Phonon calculations are performed with DFPT implemented in the Quantum Espresso package. Electron-phonon coupling matrix elements are then computed with the EPW package. We emphasize that gauge consistency of the wavefunctions in the e-ph coupling and exciton calculations is essential to obtaining the correct exciton-phonon coupling matrix elements. In our calculations, this is guaranteed by using the same wavefunction for e-ph coupling matrix elements and GW-BSE calculations. In general, the gauge consistency can also be achieved by proper gauge rotations as done in Ref. [27]. A small imaginary number with the magnitude of 75 meV is added to each $\hbar\omega$ term in the denominators in the Raman spectra calculations.

### A. IR and Raman spectrum of monolayer $MoS_2$

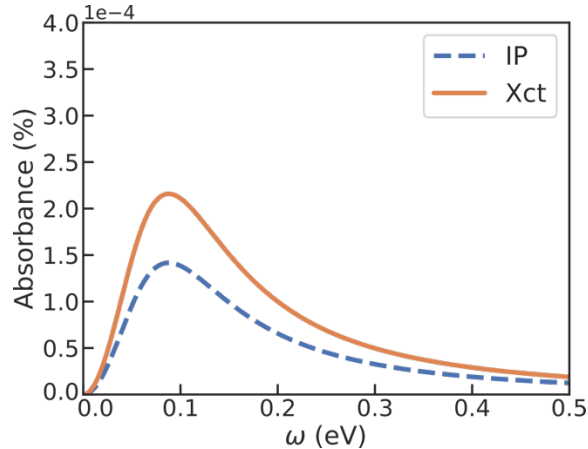

FIG. 1. (Color online) Optical absorbance due to ion motions (the IR absorption spectrum) for monolayer $MoS_2$. Results obtained in the independent-particle (IP) picture and with electron-hole interactions excitonic effects (Xct) included are shown as blue dashed and orange solid lines, respectively. The spectra are broadened with a Lorentzian function with a 75 meV broadening.

Figure 1 shows the absorbance for monolayer $MoS_2$, as computed with and without excitonic effects included. Our phonon calculations show a degenerate $E'$ zone-center phonon mode with an energy of 45.7 meV and an $A'$ mode of 48.1 meV. Among them, the $E'$ mode is IR active. Including excitonic effects does not change the peak position and spectral shape qualitatively, but the intensity is enhanced by about 50 percent. Since the absorption energy is determined by the IR-active phonon energy, to lowest order of perturbation theory, we do not expect a significant renormalization of the phonon energy from e-ph couplings.

Excitonic effects on IR spectrum can be understood by comparing the expression for the oscillator strength that goes into the expression for the dielectric susceptibility for the case with and without excitonic effects, which are given by Eq. (23) and Eq. (12), respectively. We see that both the optical transition and electron-phonon coupling matrix elements are modulated by the exciton k-space envelope functions. With the formation of exciton states, the real oscillator strength can be considered as a coherently weighted sum of the IP oscillator strengths. Similar effects also enhance the oscillator strength in the excitation of individual excitons near the band edge in the linear optical absorption spectrum. We should note that the convergence of Eq. 23 with respect the summation of all possible higher energy excitonic states is computationally demanding since in practice solving the BSE for a large number of exciton states by including large number of conduction and valence bands is a major challenge. However, excitonic effects can still be included accurately for excitonic states with energies near the band gap and the rest of the sum is kept at the IP level following the suggestion in Ref. [39].

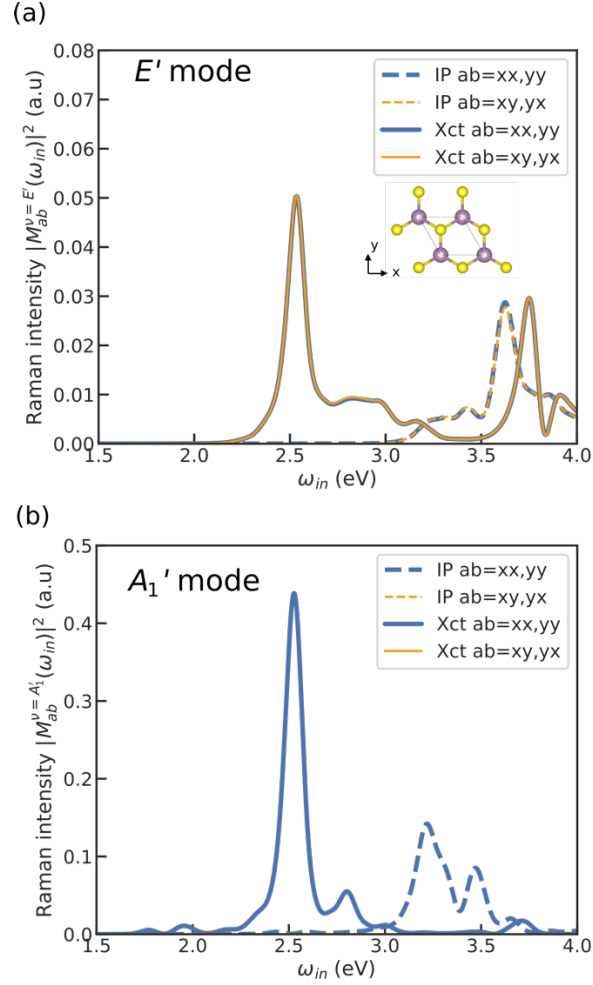

FIG. 2. (Color online) Computed absolute squares of Raman intensity tensor components in arbitrary units as a function of incident light frequency for (a) the zone-center $E'$ mode and (b) the zone-center $A_1'$ mode. Results obtained in the independent-particle (IP) picture and with excitonic effects (Xct) included are shown as dashed lines and solid lines, respectively. The xx and yy (yx and yx) components are shown with blue (orange) color. In (b) the xy and yx components are zero. Inset in (a) shows the top view of the crystal structure and the coordinate setup in our calculations, where yellow balls are chalcogen atoms and purple balls are metal atoms.

In Fig. 2 we present the absolute squares of relevant tensor components of the Raman scattering intensity as a function of incident light energy for monolayer MoS₂. We find that only $E'$ and $A_1$' modes show finite scattering intensity at both levels of theory, which agrees with previous theoretical results [1, 3]. For $E'$ modes, the xx, yy, and xy components are equal in absolute magnitudes while the amplitude of the xx and xy components differ from that of the yy component by an overall negative sign so their absolute squares fall on top of each other. The xy component of $A_1$' mode vanishes. These results also agree with the group theory analysis [1, 3]. Overall, the intensity from the $A_1'$ mode is several times larger than that of the E′ mode, which is also similar to the results reported in Ref. [13, 37].

Compared with the IP calculations, the Raman spectrum with excitonic effects included is shifted to lower energy and enhanced by several times due to the formation of bound excitons, consistent with a recent study [33]. We can identify several exciton peaks from the $A_1$' mode scattering intensity, which bear some similarity to the linear optical absorption spectrum [47]. Two small peaks at 1.8 eV and 2.0 eV in Fig. 2 (b) are identified as contributions from the A and B excitons, respectively, which correspond to 1s-like excitons formed by e-h pairs at K or K' valley. Their excitation energy difference is directly related to spin-orbit splitting of the highest occupied valence band at K and K' [47]. In Fig. 2 (a) the absence of scattering intensity at A and B exciton energy from the E′ mode was explained in Ref. [21] as a result of the nearly circular symmetry of both A and B excitons wavefunctions and the chiral nature of the E′ mode. The strongest scattering intensity for both modes is located at 2.5 eV, which is attributed to the C excitons in monolayer MoS₂[47]. We observe from numerical results that, for both modes, the dominant contributions at resonant excitations are from the double resonance terms in Eq. (25), and other contributions are at least smaller by two orders of magnitude.

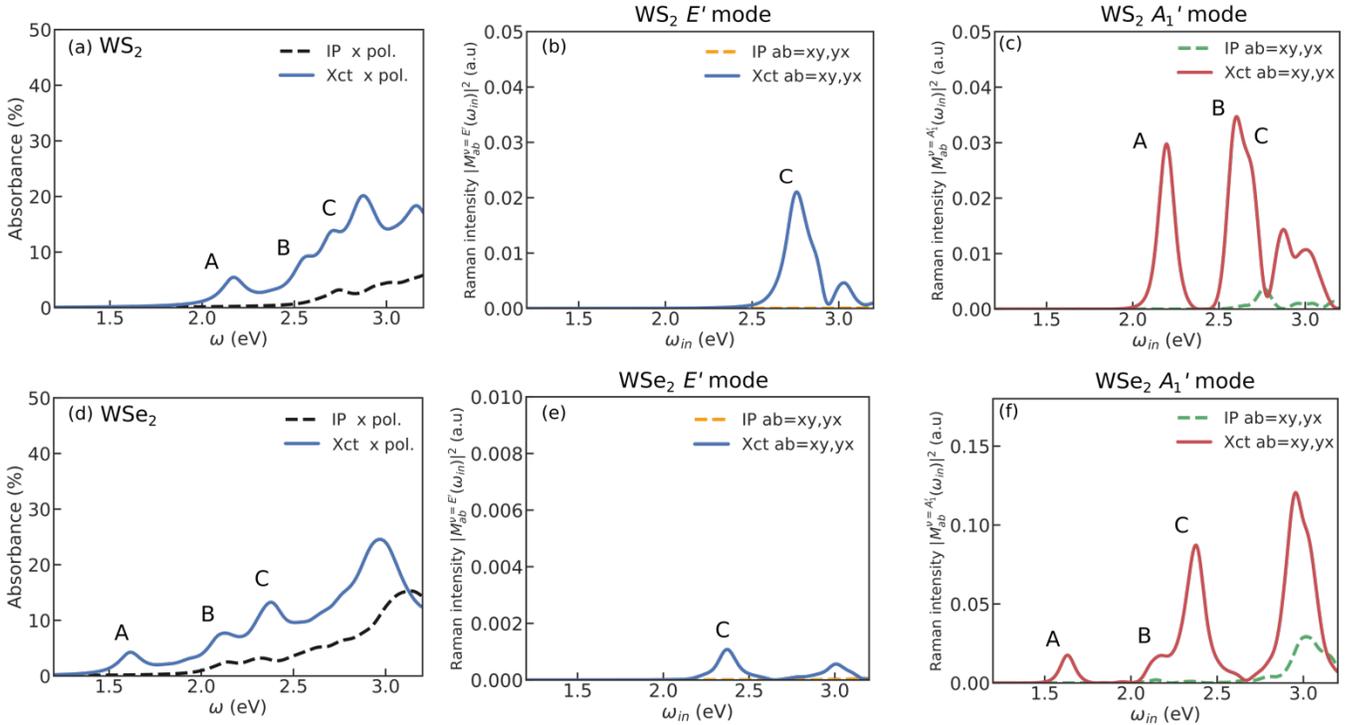

FIG. 3. (Color online) Optical absorbance (a) and (d) of monolayer WS₂ and WSe₂ as a function of incident light frequency, respectively, for linearly polarized light with polarization along the x-direction denoted as 'x pol'. Raman intensity tensor components of monolayer WS₂ (b), (c), and of monolayer WSe₂ (e), (f). Panel (b) and (e) are for the $E'$ phonon mode while (c) and (f) are for the $A_1'$ phonon mode. Results without (IP) and with excitonic (Xct) effects are shown by the dashed lines and solid lines, respectively. Resonance peaks from exciton A, B, and C of both materials are identified in each panel.

## B. Raman spectrum of monolayer WS₂ and WSe₂

Next, we study the Raman scattering spectra of monolayer WS₂ and WSe₂. From the linear optical absorption spectra shown in Fig. 3 (a) and (d), we can identify the commonly observed exciton A, B, and C peak in both materials. Similar to MoS₂, the A exciton is formed by correlated e-h pairs (interband transitions) from the band edge at the K or K' valley. The energy difference between the A and B excitons is also connected to the spin-orbit splitting of the quasiparticle bands. The computed exciton energies are in good agreement with experimental results [8, 10, 48]. We note that C excitons, on the other hand, are formed by e-h pairs not from a single K or K' valley, but they are consisting of e-h pairs from the Γ valley or along the Γ-M line, and there is more than one exciton state associated with the C exciton features. Despite the fact that all three materials (MoS₂, WS₂ and WSe₂) have similar atomic structure with D3h point group symmetry and similar band structures, they have quite different resonance Raman spectra. The $E'$ and $A_1'$ mode phonons are at frequency 44.2 and 51.8 meV for monolayer WS₂ and at frequency 29.9 and 29.8 meV for monolayer WSe₂, respectively. These modes give rise to the similar set of non-zero Raman tensor elements as monolayer MoS₂. We show the spectra of those components in Fig. 3. We observe that, as in monolayer MoS₂, excitonic effects enhance the scattering intensity and shift the spectrum to the lower energy side due to the large excitonic binding energies. Sharp features owing to excitons can be clearly identified. In Fig. 3 (b) and (e), we see prominent scattering intensity peak at the C exciton energy by the E' mode and there is no intensity at the A and B exciton energy positions, which is reminiscent of the MoS₂ case. We also attribute the vanishing intensity at the A and B exciton excitation energy as the consequence of nearly circular symmetry of their exciton envelope functions in k-space.

For the $A_1'$ mode, we can identify three exciton peaks in Fig. 3 (c) and (f) following the assignments in the absorption spectrum in Fig. 3 (a) and (b). The scattering intensity spectra of the three materials by this mode are quite different in terms of the relative intensity of the various peaks. In WSe₂, the C peak is several times stronger than the A and B peaks, while in WS₂ all three peaks have similar intensity. Similar behaviors are also seen in Ref. [8], where only a prominent C peak is observed in WSe₂ but both the A and B peaks are observed in WS₂ (although a smaller B peak than A peak in WS₂ was reported). This difference is explained as the difference in the strengths of the exciton-phonon interactions in the two materials. As we can see from Eq. (25), roughly speaking, the scattering intensity is determined by the magnitude of the excitonic dipole matrix elements, exciton-phonon coupling strengths, and whether the resonance condition in the denominator is met. We find that the dipole matrix element of the C excitons in WS₂ is only slightly larger than that in WSe₂, which can not explain the larger scattering intensity of C exciton peak in WSe₂. Therefore, a closer look of the exciton-phonon couplings is necessary to understand the difference.

In Fig. 4 we performed a detailed analysis on the exciton-phonon coupling strength (i.e., $\mathcal{G}_{ss'v}$ as defined by Eq. (24)) of the $A_1'$ mode between s=A or C excitons and s′ = all other excitons including itself. We find that the A excitons have the strongest coupling to itself or its degenerate partner state in both WS₂ and WSe₂ as shown in Fig. 4 (a) and (b), and the coupling strength is greater than that of the C excitons to other states. However, there are a few exciton states close in energy for A excitons to satisfy the resonance condition. In contrast, the C excitons in WSe₂ have more energetically close by states as shown in Fig. 4 (d). Moreover, these states couple to other nearby C excitons with similar exciton-phonon coupling strength, which explains why the C peak has a larger Raman scattering intensity compared with the A exciton or C exciton peak in WS₂.

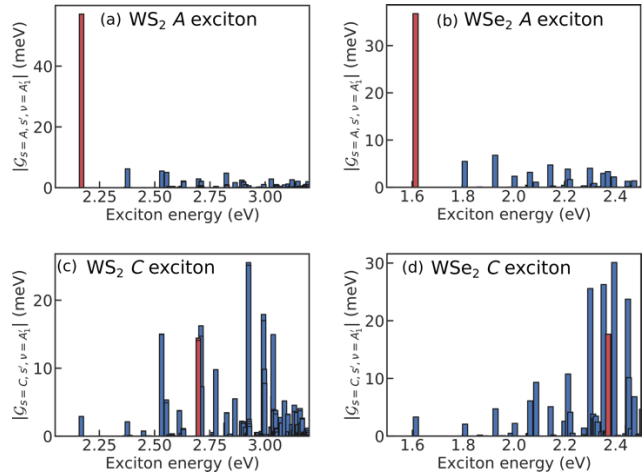

FIG. 4. (Color online) Exciton-phonon coupling matrix elements, $\mathcal{G}_{ss'v}$ between $s$ =A or C excitons and s′=all other excitons via $v = A_1'$ phonon mode in WS₂ and WSe₂. Red color highlighted the specific A or C exciton being considered.

## IV.   CONCLUSION

In conclusion, we developed a method to compute and understand IR vibrational spectrum and Raman scattering intensity including excitonic effects from first principles. Our approach is based on the TD-aGW theory with electron-phonon interactions. The derived expressions in the limit of neglecting excitonic effects reduce to previous IP results. We applied our method to monolayer MoS₂ and demonstrated that excitonic effects significantly enhanced both types of responses. We also computed resonance Raman scattering intensity in WS₂ and WS₂, and show that, despite their structural similarity, these three materials have quite different spectra, which can only be understood from the analysis of exciton-phonon coupling strength and exciton energy landscape from first principles. Going beyond lowest order perturbation theory, higher order effects can be captured by a real-time propagation of the TD-aGW equations. We expect that effects such as phonon energy renormalization, higher order Raman scatterings [49–51], excitation effects beyond the Tamm-Dancoff approximation, and temperature dependence of the spectrum [52] can be investigated through a real time-propagation approach.

### ACKNOWLEDGEMENT

This work is supported by the National Science and Technology Council of Taiwan under grant no. 110-2124-M-002-012. and by the National Science Foundation under grant no. DMR-2325410. Z.L. acknowledges support of Seed Fund from the Ershaghi Center on Energy Transition at the University of Southern California. The work is partially supported by the National Science Foundation under Grant No. OAC-2103991 in the development of interoperable software enabling the EPW and BerkeleyGW calculations with consistent gauge. We acknowledge the use of computational resources at the National Center for High-performance Computing (NCHC) of Taiwan and the Texas Advanced Computing Center (TACC) at The University of Texas at Austin. Y.H.C. thanks Michitoshi Hayashi and Shao-Yu Chen for helpful discussion.

### APPENDIX

#### 1.   Derivation of Equations 4-6

We first write down the EOM for the electron annihilation and creation operators,

$$i\hbar\frac{\partial}{\partial t}a_{mk}(t) = \epsilon_{mk}a_{mk}(t) + \sum_{l\nu}\tilde{g}_{ml\nu k}a_{lk}(t)Q_{\nu} + \sum_{j}e\boldsymbol{E}\cdot\boldsymbol{d}_{mjk}a_{jk}(t)$$

$$i\hbar\frac{\partial}{\partial t}a_{nk}^{\dagger}(t) = -\epsilon_{nk}a_{nk}^{\dagger}(t) - \sum_{l\nu}\tilde{g}_{ln\nu k}a_{lk}^{\dagger}(t)Q_{\nu} - \sum_{i}e\boldsymbol{E}\cdot\boldsymbol{d}_{ink}a_{ik}^{\dagger}(t)$$

$$(\,26\,)$$

Defining the single-electron density matrix $\rho_{mnk}(t) = \langle a_{nk}^{\dagger}(t)a_{mk}(t)\rangle$, we obtain its EOM by summing the expectation value of the product of the upper (lower) line of Eq. 26 with $a_{nk}^{\dagger}$ ($a_{mk}$). We have

$$i\hbar\frac{\partial}{\partial t}\rho_{mnk}(t) = (\epsilon_{mk} - \epsilon_{nk})\rho_{mnk}(t) + \sum_{\nu l}(\tilde{g}_{ml\nu k}\rho_{lnk}(t) - \tilde{g}_{ln\nu k}\rho_{mlk}(t))\langle Q_{\nu}(t)\rangle - \sum_{i}e\boldsymbol{E}\cdot\boldsymbol{d}_{ink}\rho_{mik}(t)$$

$$+ \sum_{j}e\boldsymbol{E}\cdot\boldsymbol{d}_{mjk}\rho_{jnk}(t).$$

$$(\,27\,)$$

We assumed a classical phonon displacement in writing down the second term in the first line. From Eq. 27 we can write down the effective time-dependent Hamiltonian, Eq. 5.

Equation 6 can be derived conveniently by rewriting the phonon Hamiltonian in terms of displacement and momentum operators as

$$H^{ph} = \frac{1}{2}\sum_{q\nu}\hbar\omega_{q\nu}(Q_{q\nu}Q_{-q\nu} + P_{q\nu}P_{-q\nu})\,,$$

where we defined the momentum operator

$$P_{q\nu} = \frac{1}{\sqrt{2}i}\left(b_{q\nu} - b_{-q\nu}^{\dagger}\right).$$

$Q_{qv}$ and $P_{qv}$ satisfy the commutation relation

$$[Q_{qv}, P_{q'\mu}] = i\delta_{q,-q'}\delta_{v\mu}.$$

The EOM of the displacement operator can be obtained from the Heisenberg equation

$$i\hbar\frac{\partial}{\partial t}Q_{q'\mu} = \left[Q_{q'\mu}, \frac{1}{2}\sum_{qv}\hbar\omega_{qv}(Q_{qv}Q_{-qv} + P_{qv}P_{-qv}) + \sum_{ijvkq}\tilde{g}_{ijv}(\boldsymbol{k},\boldsymbol{q})a_{ik+q}^\dagger a_{jk}Q_{qv}\right]$$

$$= \frac{i}{2}\sum_{qv}\hbar\omega_{qv}\delta_{\mu v}(\delta_{q',-q}P_{-qv} + \delta_{q',q}P_{qv}) = i\hbar\omega_{q'\mu}P_{q'\mu},$$

$$(28)$$

where we use $\omega_{qv} = \omega_{-qv}$ in the last line. The EOM of the momentum operator reads

$$i\hbar\frac{\partial}{\partial t}P_{q'\mu} = \left[P_{q'\mu}, \frac{1}{2}\sum_{qv}\hbar\omega_{qv}(Q_{qv}Q_{-qv} + P_{qv}P_{-qv}) + \sum_{ijvkq}\tilde{g}_{ijv}(\boldsymbol{k},\boldsymbol{q})a_{ik+q}^\dagger a_{jk}Q_{qv}\right]$$

$$= -i\hbar\omega_{q'\mu}Q_{q'\mu} - i\sum_{ijvkq}\tilde{g}_{ijv}(\boldsymbol{k},\boldsymbol{q})a_{ik+q}^\dagger a_{jk}\delta_{q'-q}\delta_{\mu v} = -i\hbar\omega_{q'\mu}Q_{q'\mu} - i\sum_{ijk}\tilde{g}_{ij\mu}(\boldsymbol{k},-\boldsymbol{q'})a_{ik-q'}^\dagger a_{jk}.$$

$$(29)$$

Eq. 6 is obtained by taking the time-derivative of Eq. 28 and replacing the time-derivative of the momentum operator with Eq. 29. We arrive at

$$i\hbar\frac{\partial^2}{\partial t^2}Q_{q\mu} = -i\hbar\omega_{q\mu}^2 Q_{q\mu} - i\omega_{q\mu}\sum_{ijk}\tilde{g}_{ij\mu}(\boldsymbol{k},-\boldsymbol{q})a_{ik-q}^\dagger a_{jk}.$$

### 2. Equivalence to the previous derivation of IR spectrum

We compare our results, Eq. 11, with Eq. 27 in Ref. [24] which is

$$\chi^{\alpha\beta}(\omega) = \frac{1}{V}\sum_s \frac{f_{s\alpha}(\omega)f_{s\beta}(\omega)}{\omega_s^2 - \left(\omega + \frac{i\gamma}{2}\right)^2},$$

$$(30)$$

where $f_{s\alpha}(\omega)$ is the oscillator strength, defined as

$$f_{s\alpha}(\omega) \equiv e\sum_{l\mu}Z_{\alpha\mu}^l(\omega)\frac{e_{s,\mu}^l}{\sqrt{M_l}} = \frac{2e}{N_p}\sum_{l\mu}\sum_{kcv}\frac{i\hbar}{\hbar\omega - \epsilon_{cvk}}\frac{v_{vck}^\alpha}{\epsilon_{cvk}}\left\langle u_{ck}\left|\frac{\partial V^{KS}}{\partial u_\mu^l}\right|u_{vk}\right\rangle\frac{e_{s,\mu}^l}{\sqrt{M_l}} = \frac{2e}{N_p}\sum_{kcv}\frac{-r_{vck}^\alpha g_{cvs}}{\hbar\omega - \epsilon_{cvk}}\left(\frac{2\omega_s}{\hbar}\right)^{\frac{1}{2}},$$

$$(31)$$

where $Z_{\alpha\mu}^l(\omega)$ is the dynamical Born effective charge, $|u_{ck}\rangle$ is the periodic part of the Bloch state with the band index c, $v_{ijk} = \frac{i\epsilon_{ijk}r_{ijk}}{\hbar}$ and we use the definition of electron-phonon coupling matrix elements [34],

$$g_{ijs}(\boldsymbol{k},\boldsymbol{q}) = \sum_{\kappa\alpha}\left(\frac{\hbar}{2\omega_{qs}M_\kappa}\right)^{\frac{1}{2}}e_{s,\alpha}^\kappa\sum_p e^{-i\boldsymbol{q}\cdot(\boldsymbol{r}-\boldsymbol{R}_p)}\left\langle u_{ik}\left|\frac{\partial V^{KS}}{\partial u_\alpha^\kappa}\right|u_{jk}\right\rangle,$$

$$(32)$$

in the last equality of Eq. 31. Inserting $f_{s\alpha}(\omega)$ into the equation, we have

$$\chi^{\alpha\beta}(\omega) = \sum_s \frac{C\omega_s}{\omega_s^2 - \left(\omega + \frac{i\gamma}{2}\right)^2}\left(\sum_{kcv}\frac{r_{vck}^\alpha g_{cvs}}{\hbar\omega - \epsilon_{cvk}}\right)\left(\sum_{kcv}\frac{r_{vck}^\beta g_{cvs}}{\hbar\omega - \epsilon_{cvk}}\right),$$

where $C = 8e^2/N_p^2 V\hbar$. The factor $8/N_p$ accounts for the spin degeneracy and the definition of $\tilde{g}$.